\documentclass[aps,prb,twocolumn,showpacs,preprintnumbers,amsmath,amsfonts,amssymb,floatfix,aps,superscriptaddress,longbibliography]{revtex4-2}
\usepackage[english]{babel}

\usepackage{graphicx}
\usepackage{dcolumn}
\usepackage{bm}
\usepackage[utf8]{inputenc}
\usepackage[T1]{fontenc}
\usepackage{etoolbox}
\usepackage{verbatim}
\usepackage{hyperref}
\hypersetup{
    colorlinks=true,
    linkcolor=blue,
    filecolor=magenta,      
    urlcolor=blue,
    citecolor=blue,
}
\usepackage{bbold}
\usepackage{mathrsfs}
\usepackage{amsmath,amsfonts,amssymb}
\usepackage{braket}
\usepackage{url}
\usepackage[dvipsnames]{xcolor}
\usepackage[normalem]{ulem}

\begin{document}
\title{Supercurrents in Josephson junctions with chiral molecular potentials}

\author{Oleg Kuliashov}

\affiliation{Applied Physics Department, The Hebrew University of Jerusalem, Jerusalem, Israel}

\author{Alberto Cappellaro}

\affiliation{Dipartimento di Fisica ed Astronomia \textit{"G. Galilei"}, Universit\'a degli Studi di Padova, via Marzolo 8, 35131 Padova, Italy}

\affiliation{National Institute of Nuclear Physics (INFN), Padova Section, 
    Via Marzolo 8, 35131 Padova, Italy}
    
\affiliation{Institute of Science and Technology Austria (ISTA), Am Campus 1,
3400 Klosterneuburg, Austria}

\author{Oded Millo}
\affiliation{The Racah Institute of Physics, The Hebrew University of Jerusalem, 91904 Jerusalem, Israel}

\author{Yossi Paltiel}
\affiliation{Applied Physics Department, The Hebrew University of Jerusalem, Jerusalem, Israel}

\author{Mikhail Lemeshko}
\affiliation{Institute of Science and Technology Austria (ISTA), Am Campus 1,
3400 Klosterneuburg, Austria}

\author{Ragheed Alhyder}
\email{ragheed.alhyder@ista.ac.at}
\affiliation{Institute of Science and Technology Austria (ISTA), Am Campus 1,
3400 Klosterneuburg, Austria}

\begin{abstract}
The influence of chiral molecular potentials on phase-coherent transport in superconducting Josephson junctions is investigated. Within a Bogoliubov--de Gennes tight-binding framework, a superconducting--normal--superconducting (SNS) junction functionalized by adsorbed chiral molecules is modeled, where electrostatic gradients generated by the molecules induce spin--orbit coupling in the normal region. The equilibrium charge current--phase relation is found to remain largely insensitive to molecular chirality in symmetric, zero-field configurations. In contrast, the spin-polarized equilibrium current exhibits a pronounced chirality-dependent response, with opposite enantiomers producing distinct and anisotropic spin-polarized Josephson currents. The resulting handedness contrast can be enhanced through control parameters such as molecular orientation and the strength of the induced spin--orbit coupling. The temperature dependence of these currents further shows that the chirality-dependent signatures persist across a range of temperatures well below the superconducting critical temperature. These results identify Josephson interferometry as a phase-sensitive setting for probing chirality-dependent spin structure and highlight spin-polarized superconducting transport as a controlled route toward integrating chiral molecular functionality into superconducting spintronic devices.
\end{abstract}
\maketitle
\section{Introduction}
Chirality is a geometric property of matter that has far reaching consequences across chemistry, biology, and condensed matter physics, because it distinguishes structures that are related by spatial inversion but cannot be continuously deformed into one another. In molecular settings, opposite enantiomers have the same energy but coexist with strikingly different functional behavior, which is why chirality is central to stereospecific chemistry and to the emergence of biological homochirality in living systems \cite{barron-1986,saito-2013,blackmond-2019}. At the same time, the role chirality plays in the electronic and magnetic properties of materials is increasingly investigated, since it can imprint handedness onto transport and response functions once spin and orbital degrees of freedom are involved~\cite{Aiello-2022}. Exploiting these effects to establish scalable ways to interrogate chirality through robust physical observables remains an active challenge, complementing powerful spectroscopic approaches that access chirality through light matter interactions or enantiomer specific rotational responses \cite{doyle-2013}.

A particularly influential development in this direction is chirality induced spin selectivity (CISS), where charge transport through chiral systems becomes correlated with spin polarization \cite{naaman-1999,xie-2011,gutierrez-2012,naaman_spintronics_2015,waldeck-2019,naaman-2020}. Experiments have connected such spin selectivity to functionality ranging from chiral molecular spin filters and enantiospecific interactions with magnetic substrates to magnetization control driven by adsorbed chiral layers \cite{banerjee-ghosh-2019,BenDor2017,xie-2011,gutierrez-2012,guo-2012-1,guo-2012-2,naaman_spintronics_2015,varela-2016,aragones-2017,alam-2017,lu-2019,sessoli-2019,waldeck-2019,geyer-2019,vanwees-2019,volosniev-2020,vanwees-2020,naaman-2020,volosniev-2021,kulkarni-2020,evers_theory_2022,ozturk-2023,wasielewski-2023,el-naggar-2023,menichetti-2025,kapon2025,Moharana2025}. Theoretical progress has clarified multiple microscopic routes that can generate sizable spin selectivity while also highlighting open questions regarding quantitative modeling, the role of environment and dissipation, and the precise symmetry requirements for chirality specific signals in realistic devices \cite{ Fransson2019, Ghazaryan2020a,evers_theory_2022,Hedegrd2023, alhyder-2023,alhyder2025b,DiVentra2025}. This combination of broad relevance and unresolved microscopic structure motivates the search for device platforms where chirality dependent physics can be amplified and read out reliably.

Josephson junctions provide a natural arena for such an approach because they convert subtle changes in quantum phase accumulation into macroscopic currents that can be measured with exquisite precision \cite{josephson-1962}. Their equilibrium supercurrent is carried by Andreev processes and bound states whose phase dispersion is controlled by scattering phases and spin-dependent structure in their normal region \cite{andreev-1964,btk-1982,beenakker-1991,golubov-2004}. In practice, Josephson devices form the backbone of superconducting electronics, serving both as ultrasensitive interferometers in superconducting quantum interference device (SQUID) based metrology and as engineered nonlinear elements in leading quantum information architectures \cite{clarke-2008,devoret-2013,weldin-2017,kjaergaard-2020,krantz-2019}. The same phase coherence that enables these applications also makes Josephson junctions exceptionally responsive to spin--orbit coupling and spin-active scattering, including regimes where symmetry permits anomalous phase shifts or unconventional current components \cite{rashba-1984,manchon-2015,bercioux-2015,buzdin-2008}. More broadly, recent works have demonstrated that structural chirality
can strongly influence quantum transport phenomena, for example leading to
nonreciprocal superconducting currents in Josephson junctions formed from
opposite-handed crystals~\cite{Debnath2024, Debnath2025, orban2026}, and that chiral spin-orbit coupling can by itself produce an unconventional Josephson diode response~\cite{Costa2025}.

These observations motivate a synergy between CISS-based molecular physics and superconducting interferometry, in which a chiral molecular texture acts as a spatially structured spin orbit and potential landscape that reshapes Andreev transport. Experimental indications that chiral molecular layers can couple to superconducting states and induce magnetic related signatures strengthen this perspective and motivate superconducting circuits as a practical readout modality for molecular chirality \cite{Alpern-2019,Alpern2021}. In parallel, it was shown that chiral molecular potentials can generate symmetry controlled chirality dependent transport responses on magnetized surfaces, with a crucial dependence on molecular orientation and electrostatic gradients \cite{alhyder2025b}. These results frame chiral molecular junction functionalization as a tunable route to spin active superconducting junctions whose response can be engineered without requiring microscopic state resolution.

Here, we develop a Bogoliubov--de Gennes (BdG) transport framework for SNS Josephson junctions functionalized by chiral molecular textures, and we show how chirality enters superconducting observables through spin-dependent scattering phases accumulated in the normal region. We compute both the charge current--phase relation and equilibrium spin-polarized currents and demonstrate that chirality contrast is naturally enhanced by control knobs that activate interferometric sensitivity. These include orbital magnetic field patterns and the strength of molecule-induced spin--orbit coupling. We further analyze the temperature dependence of these currents, which identifies the regime where chirality-dependent signatures persist despite thermal smearing. Taken together, these results position Josephson interferometry as a phase-sensitive probe of chirality-dependent spin structure in electronic nanodevices, while simultaneously providing a controlled setting to investigate the interplay of chiral spin physics and superconducting phase coherence.


\section{Model}

We model an SNS Josephson junction on a two-dimensional lattice with lattice constant $a$, with length $L$ and width $W$. Two superconducting leads occupy the regions $x<x_\ell$ (left lead) and $x\ge x_r$ (right lead), while the normal component spans $x_\ell\le x < x_r$. The phase bias between the superconductors is denoted by $\phi$.

We work in the Nambu spinor basis
$
\hat{\Psi}_i
=
\big(\hat{c}_{i\uparrow},\,\hat{c}_{i\downarrow},\,\hat{c}^\dagger_{i\uparrow},\,\hat{c}^\dagger_{i\downarrow}\big)^{\mathsf T},
$
and use Pauli matrices $\boldsymbol{\tau}=(\tau_x,\tau_y,\tau_z)$ in particle--hole (Nambu) space and $\boldsymbol{\sigma}=(\sigma_x,\sigma_y,\sigma_z)$ in spin space, with identities $\tau_0$ and $\sigma_0$.

The mean-field Bogoliubov--de~Gennes Hamiltonian is written as
\begin{equation}
\hat{H}_{\mathrm{BdG}}(\phi)
=
\frac12\sum_{ij}\hat{\Psi}_i^\dagger
\begin{pmatrix}
h_{ij} & \Delta_{ij}(\phi)\\
\Delta_{ij}^\dagger(\phi) & -h_{ij}^\ast
\end{pmatrix}
\hat{\Psi}_j,
\label{eq:BdG_general}
\end{equation}
where $i,j$ label lattice sites and $h_{ij}$ ($\Delta_{ij}$) are $2\times 2$ matrices in spin space. 
The normal-state Hamiltonian $h$ is decomposed as
$h = h_{\mathrm{kin}} + h_{Z}
$,
where $h_{\mathrm{kin}}$ is the kinetic energy including the chemical potential
$
(h_{\mathrm{kin}})_{ij}=(4t-\mu)\,\sigma_0\,\delta_{ij}-t\,\sigma_0\,\delta_{\langle ij\rangle},
$
with nearest-neighbor hopping $t>0$, the onsite term $4t$ is the standard shift that places the band minimum at $-\mu$, $\delta_{\langle ij\rangle}=1$ for nearest neighbors and $0$ otherwise. A magnetic field along $\hat{z}$ is included through a Zeeman term
$(h_Z)_{ij}=h_z\,\sigma_z\,\delta_{ij}.$

Superconductivity is imposed through an onsite spin-singlet pairing potential in the lead regions with a phase difference $\phi$ across the junction:
\begin{equation}
\Delta_{ij}(\phi)
=
\Delta_i\,e^{i\phi_i}\,(i \sigma_y)\,\delta_{ij},
\label{eq:Delta_onsite}
\end{equation}
with the piecewise-constant phase profile $\phi_i=-\phi/2$ for $x_i<x_\ell$, $\phi_i=0$ for $x_\ell\le x_i<x_r$, and $\phi_i=+\phi/2$ for $x_i\ge x_r$.
The pair potential is imposed as an external proximity field, neglecting inverse-proximity feedback effects on the superconducting order parameter.
Orbital effects are included through a Peierls phase on nearest-neighbor hoppings. Using a Landau-gauge in which only hoppings along $\hat{y}$ acquire a phase, and only within the normal region $x_\ell\le x < x_r$, we write the hopping term $ -t\,e^{i\theta_{ij}\tau_z},$
with $
\theta_{ij} = B\,\kappa\, x_j\,(y_i-y_j),
$
The explicit $\tau_z$ again ensures opposite coupling for particles and holes. When orbital effects are not required one sets $B=0$.

\begin{figure}[t]
\centering
\includegraphics[width=\columnwidth]{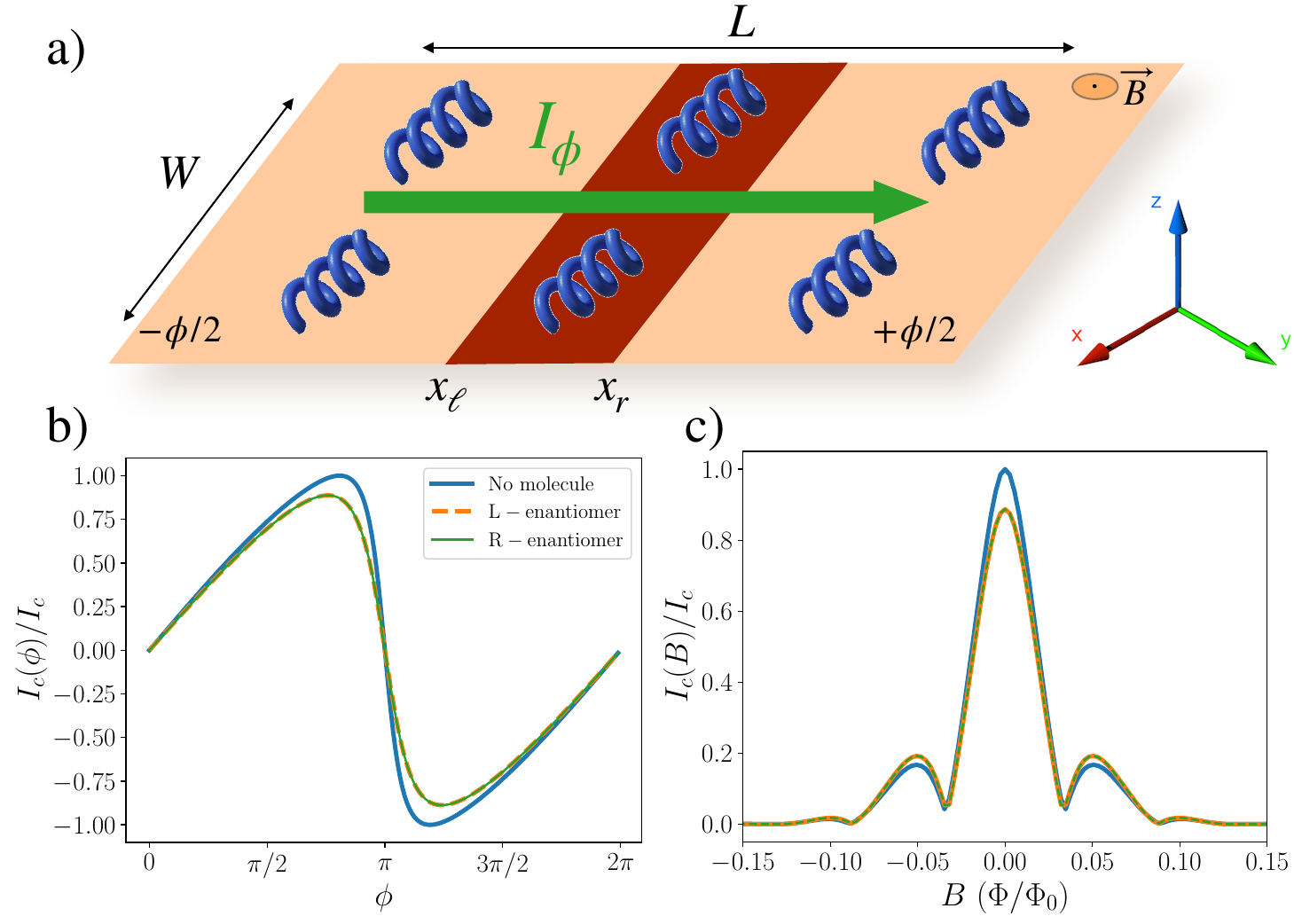}
\caption{
\textbf{(a)} Schematic of the Josephson junction considered throughout this work. Two superconducting leads with order-parameter phases $\mp\phi/2$ are connected through a normal weak link (dark region) with adsorbed chiral molecular structures on top. The equilibrium supercurrent $I_\phi$ flows along the transport direction in response to the gauge-invariant phase bias $\phi$. An out-of-plane magnetic field $B\hat{\mathbf{z}}$ threads flux through the junction area and produces interference of transverse supercurrent paths. 
\textbf{(b)} Current--phase relation at zero field, $I_c(\phi)\equiv I(\phi,B{=}0)$. While the molecular texture visibly renormalizes the overall Josephson response relative to the no-molecule reference, the two enantiomers yield overlapping charge supercurrents in this parameter set, which is consistent with chirality entering predominantly through the spin-dependent structure of the Andreev states. Robust chirality readout therefore requires observables that couple directly to the spin sector or additional symmetry breaking.
\textbf{(c)} Fraunhofer response: critical current $I_c(B)=\max_{\phi}|I(\phi,B)|$ normalized to its zero-field value, comparing the reference junction without molecules (blue) to junctions functionalized by left- and right-handed enantiomers (orange and green). 
}
    \label{fig:cpr}
\end{figure}
The molecule contributes a scalar onsite term to the BdG Hamiltonian,
\begin{equation}
\hat{H}^{(0)}_{\mathrm{mol}}
=
\sum_i \hat{\Psi}_i^\dagger \Big[ V_{\mathrm{mol}}(\mathbf{r}_i)\,\tau_z\otimes \sigma_0 \Big]\hat{\Psi}_i,
\label{eq:mol_scalar}
\end{equation}
where $V_{\mathrm{mol}}(\mathbf{r})$ is the electrostatic potential generated by the molecule, evaluated at lattice positions $\mathbf{r}_i=(x_i,y_i)$. The explicit $\tau_z$ structure ensures that electrons and holes experience opposite scattering potentials.
We model each adsorbed molecule by an electrostatic potential~\cite{Ghazaryan2020a,alhyder2025b}
\begin{equation}
V_{\mathrm{mol}}(\mathbf r)
=
\bm{\mathcal{E}}\cdot \bm r'\,
\exp\!\left[
-\frac{1}{2}\left(\frac{x'^2}{\ell_x^2}+\frac{y'^2}{\ell_y^2}+\frac{z'^2}{\ell_z^2}\right)
\right],
\label{eq:Vmol_codeform}
\end{equation}
where $\bm r'=(x',y',z')$ are coordinates in the molecular frame, obtained by rotating the molecule about the $x$ axis by an angle $\theta$, i.e.\ $\bm r'=R_x(\theta)\,(\mathbf r-\mathbf r_m)$ with $R_x(\theta)$ the standard right-handed rotation matrix, and $\mathbf r_m=(x_m,y_m,0)$ the molecular center. 
The parameters $\ell_{x,y,z}$ set the anisotropic spatial extent of the potential. The linear prefactor is controlled by an effective internal field $\bm{\mathcal{E}}=(\mathcal{E}_x,\mathcal{E}_y,\mathcal{E}_z)$, which we parametrize in terms of an effective molecular dipole moment $\bm\mu=(\mu_x,\mu_y,\mu_z)$ as
$
\mathcal{E}_\nu=8e\,\mu_\nu/\ell_\nu^{3},
$ with $\nu\in\{x,y,z\}$, and $e$ the electron charge in our dimensionless convention. The explicit molecular parameters used in the simulations are summarized in Appendix~\ref{appA:mol_potential}.

In the present model, opposite molecular chiralities are implemented by a single discrete parameter $\chi=\pm 1$ multiplying the $y$-component of the molecular moment $\mu_y$, while all other parameters are held fixed.
This choice corresponds to a mirror operation with respect to the $xz$ plane in the molecular parametrization (i.e., $y\mapsto -y$ in the molecular frame), which reverses the handedness of the potential.
Independently, the continuous angle $\theta$ controls the molecular orientation relative to the junction plane. At fixed laboratory orientation, changing $\chi$ compares two enantiomeric molecular textures placed in the same junction. The exact mirror-related device is obtained only when the molecular orientation is reversed together with the handedness, so that the mirror $M_{xz}$ relates the configuration $(\chi,\theta)$ to $(-\chi,-\theta)$. This distinction is important for interpreting the angular dependence, because symmetry constrains pairs of configurations related by the full mirror rather than the fixed-$\theta$ scan point by point. We use this relation to explain the spin-current signatures in Sec.~\ref{subsec:theta-scan}.

Spin--orbit effects arise from the inhomogeneous molecular potential via a Pauli (Rashba-like) coupling~\cite{rashba-1984}, implemented on the lattice as spin-dependent nearest-neighbor hoppings
\begin{equation}
\hat{H}_{\mathrm{mol}}^{(\mathrm{SO})}
=
\sum_{\langle ij\rangle}
\hat{\Psi}_i^\dagger
\Big[
-i\,\alpha_{\mathrm{SO}}\,
\big[\bm{\mathcal E}(\mathbf r_i)\times \hat{\mathbf d}_{ij}\big]\cdot(\tau_z\otimes \boldsymbol{\sigma})
\Big]\hat{\Psi}_j
+ \mathrm{H.c.},
\label{eq:mol_soc_compact}
\end{equation}
where $\alpha_{\mathrm{SO}}$ is the spin-orbit parameter indicating the strength of the coupling, $\bm{\mathcal E}(\mathbf r)\equiv-\nabla V_{\mathrm{mol}}(\mathbf r)$ is the local effective electric field, $\hat{\mathbf d}_{ij}=(\mathbf r_i-\mathbf r_j)/|\mathbf r_i-\mathbf r_j|$ for nearest neighbors, and $\tau_z$ enforces the opposite sign in the hole sector.

We work in dimensionless tight-binding units, setting the lattice spacing and nearest-neighbor hopping to unity ($a\equiv1$, $t\equiv1$). All energies in the Hamiltonian are therefore measured in units of $t$, and temperature is measured in units of $t/k_B$ (equivalently $k_B\equiv1$). Orbital magnetic effects enter through a Peierls phase on hoppings with the dimensionless choice $\theta_{ij}=B\,x_j(y_i-y_j)$ (i.e.\ Peierls prefactor set to unity). Physical units can be restored by reinstating $\hbar$ and $|e|$ in $\theta_{ij}=(|e|/\hbar)\int_{\mathbf r_j}^{\mathbf r_i}\mathbf A\cdot d\boldsymbol{\ell}$ and identifying $t=\hbar^2/(2m^\ast a^2)$, so that all dimensionless energies map to physical scales by multiplication with $t$.

\section{Results}\label{sec:results}
We perform numerical simulations on the Hamiltonian described previously by exact diagonalization of the BdG Hamiltonian in Eq.~\eqref{eq:BdG_general} for a given set of parameters, which yields the quasiparticle spectrum $\{E_n(\phi,B)\}$ and eigenstates $\{|n\rangle\}$ as a function of superconducting phase bias $\phi$ and orbital magnetic field $B$. We then compute the equilibrium charge and spin supercurrents from these spectral data with the help of the Kwant package~\cite{Groth-2014,Kloss-2021} as detailed below. The resulting currents are analyzed as a function of control parameters such as the effective spin--orbit coupling (SOC) strength $\alpha$, molecular orientation $\theta$, and temperature $T$.

\subsection{Junction geometry and computational protocol}\label{subsec:setup-protocol}

Figure~\ref{fig:cpr}(a) summarizes the superconducting--normal--superconducting (SNS) junction geometry studied in this work. Two superconductors with order-parameter phases $\mp\phi/2$ are coupled through a normal link, with chiral molecular texture adsorbed on the surface of the junction. The handedness of this texture is encoded by a chirality index $\chi=\pm1$ (corresponding to left- and right-handed enantiomers), and it generates a spin-dependent region through an effective spin--orbit coupling tied to the electrostatic gradients produced by the molecule. An external magnetic field $B\hat{\mathbf z}$ can generate flux through the junction area, enabling interference of transverse supercurrent trajectories and thereby providing an experimentally direct probe of phase-coherent transport.

For a given set of microscopic parameters, we compute the equilibrium Josephson current from the BdG quasiparticle spectrum $\{E_n(\phi,B)\}$ via the standard thermodynamic expression
\begin{equation}
I(\phi,B)=\frac{2e}{\hbar}\sum_{n>0}\tanh\!\left(\frac{E_n(\phi,B)}{2T}\right)\,\frac{\partial E_n(\phi,B)}{\partial \phi},
\label{eq:Josephson_from_spectrum_final}
\end{equation}
where the sum runs over positive BdG energies. This formulation makes explicit how molecule-induced spin-active scattering reshapes the phase sensitivity of the spectrum and hence the supercurrent response.

Figure~\ref{fig:cpr}(b) displays the equilibrium current--phase relation at $B=0$ for three configurations: a reference junction without molecules and junctions functionalized by left- and right-handed enantiomers. In the absence of molecules, the current exhibits the characteristic periodic behavior of an SNS junction. Introducing the molecular texture renormalizes the Josephson response, modifying the overall critical current and slightly reshaping the current–phase relation, reflecting the fact that the molecule modifies the effective transparency and the spin-dependent phase accumulation of Andreev quasiparticles. The phase dispersion of the low-energy Andreev bound states underlying the Josephson response is illustrated in Appendix~\ref{appD:abs_spectrum}.

In addition, we show in figure~\ref{fig:cpr}(c) the Fraunhofer response, i.e., the orbital-field dependence of the critical current
$
I_c(B)=\max_{\phi}\,|I(\phi,B)|,
$
normalized to its zero-field value. The out-of-plane field $B\hat{\mathbf z}$ induces a position-dependent gauge-invariant phase accumulation across the junction width, leading to interference between supercurrent contributions carried by different transverse trajectories. The resulting $I_c(B)$ exhibits the familiar lobe structure characteristic of Josephson interferometry, providing a sensitive and experimentally standard diagnostic of phase-coherent transport through the weak link.
Relative to the no-molecule reference, the molecular texture again produces a systematic suppression and reshaping of the envelope.

A notable feature in both figures is that the charge current for the two enantiomers is nearly indistinguishable in this parameter set. This behavior is consistent with chirality entering predominantly through the spin-dependent structure of the quasiparticle states in the normal region, while the charge supercurrent at zero field remains largely constrained by equilibrium symmetries. In particular, in the absence of an explicit time-reversal-breaking perturbation, opposite enantiomers yield similar energies and therefore similar charge currents. This constraint is exact. The handedness reversal is the mirror $M_{xz}$, and the charge current along $\hat{x}$ lies in the mirror plane, so it is even under $M_{xz}$ and enantiomers related by the full mirror carry identical charge current at zero field. The residual splitting of the order of $10^{-4}$ in Figs.~\ref{fig:alpha}(a) and \ref{fig:theta}(a) is allowed by symmetry, because reversing $\chi$ at fixed orientation is not the full mirror for a tilted molecule, and it is the small $\theta$-odd component of the charge current rather than a charge diode, which at zero field is forbidden by time reversal, $I(-\phi)=-I(\phi)$. The spin sector carries the handedness information and follows from the same mirror, analyzed in Sec.~\ref{subsec:theta-scan}. This motivates complementing charge-transport interferometry with probes that couple more directly to the spin sector (e.g., spin-polarized Josephson observables) and, crucially, exploring regimes with enhanced spin mixing and explicit time-reversal breaking, where enantiomer-dependent corrections to transport become parametrically larger.
\subsection{Spin-polarized Josephson transport and tuning by spin--orbit coupling}\label{subsec:alpha-scan}
\begin{figure}
    \centering
    \includegraphics[width=\columnwidth]{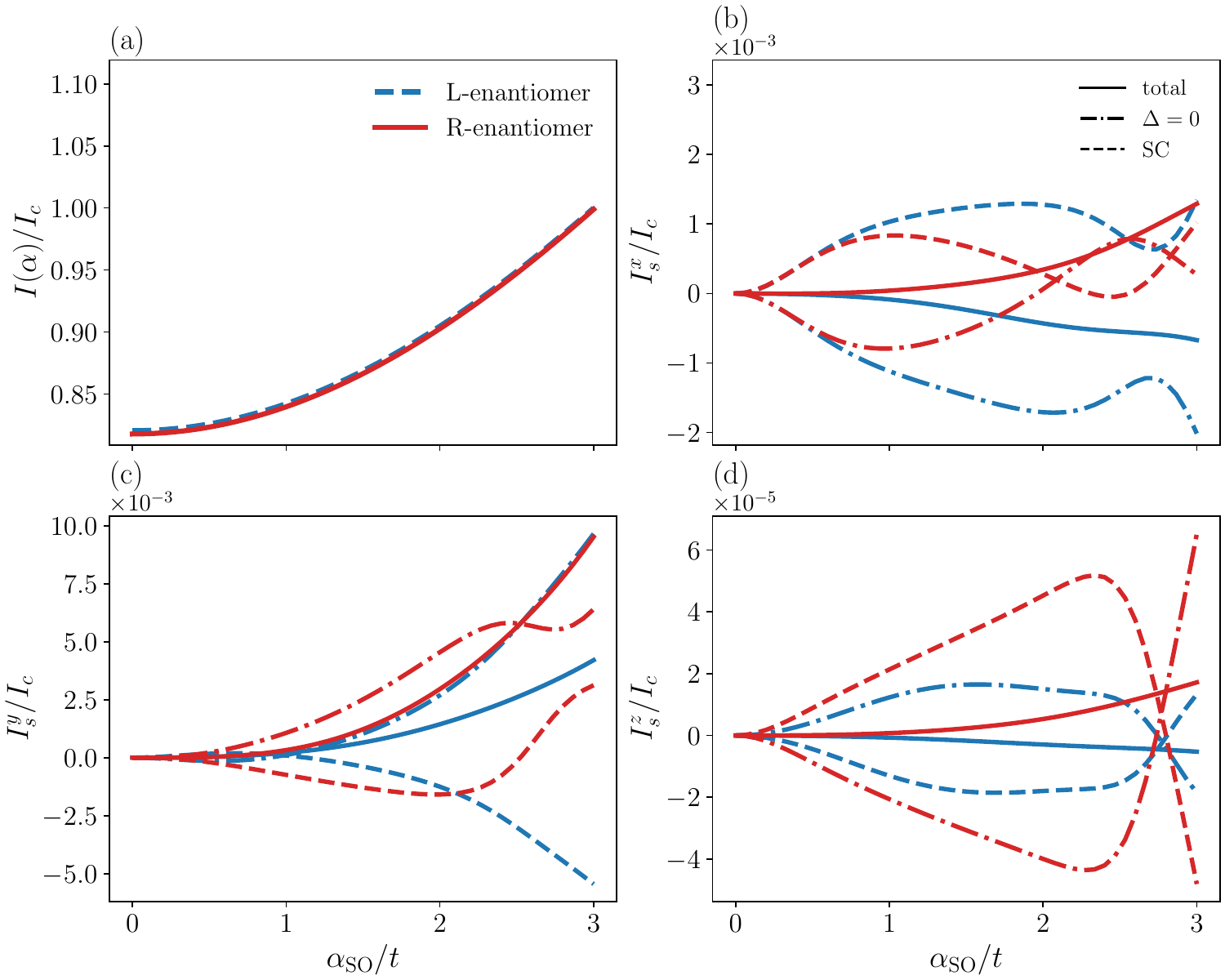}
    \caption{
Dependence of charge and spin currents on the effective spin--orbit coupling strength 
$\alpha_{\mathrm{SO}}$. 
(a) The normalized charge current at fixed phase bias varies only weakly with handedness, indicating that 
the primary molecular effect in the charge sector is a renormalization of the junction transparency. 
(b)--(d) In contrast, finite spin currents emerge in the presence of the molecular potential and grow
with increasing $\alpha_{\mathrm{SO}}$, with distinct chirality-dependent responses for the two enantiomers.
In the spin panels the total current (solid) is separated into the normal-state contribution at $\Delta=0$ (dash-dotted) and the superconductivity-induced part (dashed), for the left-handed (blue) and right-handed (red) enantiomers. All currents are normalized to the charge critical current $I_c$.
Parameters: $W=16$, $L=46$, $\mu=0.5$, $\Delta_0=0.2$, $h_z=0$, $\theta=\pi/4$, $\phi=\pi/2$, and $B=0$.
}
    \label{fig:alpha}
\end{figure}
Figure~\ref{fig:alpha} summarizes how the charge and spin-polarized equilibrium currents evolve as the effective spin--orbit coupling strength $\alpha$ associated with the molecular electrostatic gradients is tuned. The charge current is evaluated from Eq.~\eqref{eq:Josephson_from_spectrum_final}. The spin-current components are computed from the equilibrium expectation value of the corresponding conventional bond-current operators across a cut through the junction: denoting by $\hat{J}^{a}$ the spin-polarized bond-current operator for spin polarization $a\in\{x,y,z\}$, we evaluate
\begin{equation}
I_s^{\nu}(\phi)=\frac{1}{2}\sum_{n>0}\tanh\!\left(\frac{E_n(\phi)}{2T}\right)\,\langle n(\phi)|\hat{J}^{\nu}|n(\phi)\rangle,
\label{eq:spin-current}
\end{equation}
where $|n(\phi)\rangle$ are BdG eigenstates with positive energies and the prefactor avoids double counting associated with the Nambu representation. In systems with spin--orbit coupling this is the conventional spin-current definition, and spin is not a locally conserved quantity because spin torque terms are present. We therefore interpret $I_s^\nu$ as a spin-polarized equilibrium bond-current expectation value. To separate the superconductivity-induced response from the equilibrium spin current already present in the normal state, we define the superconducting part
\begin{equation}
I^{\nu}_{s,\mathrm{SC}} = I_s^\nu - I_s^\nu(\Delta=0),
\label{eq:sc-spin-current}
\end{equation}
where $I_s^\nu(\Delta=0)$ is evaluated in the same structure with the pairing set to zero. The normal-state term $I_s^\nu(\Delta=0)$ is the equilibrium spin current of the broken-inversion spin--orbit-coupled system, essentially temperature independent and finite above $T_c$~\cite{yi2006a,sun2007}, whereas $I^{\nu}_{s,\mathrm{SC}}$ originates from the chirality-dependent modification of the Andreev spectrum, scales with the superconducting gap, and is the Josephson-specific part of the chirality signal. Both contributions are shown together with the total current in Figs.~\ref{fig:alpha}, \ref{fig:theta}, and \ref{fig:T}. A derivation of Eq.~\eqref{eq:spin-current} within the Bogoliubov–de Gennes formalism is outlined in Appendix~\ref{appB:current_operators}.

Two robust trends are apparent in Fig.~\ref{fig:alpha}. First, the charge response shows only a weak dependence on chirality for the parameters considered: while introducing the molecular texture produces a clear, systematic renormalization of the charge critical current relative to the no-molecule reference as $\alpha$ is increased, the left- and right-handed enantiomers remain nearly indistinguishable on the scale of Fig.~\ref{fig:alpha}(a). This indicates that, here, the leading charge-current effect of the molecules is largely scattering represented in an overall change in effective transparency and phase dispersion.
Second, the spin-polarized currents provide a much more direct readout of molecular handedness, reflecting the spin-dependent phase accumulated by quasiparticles while propagating through the chiral molecular potential, which depends on the molecular orientation angle $\theta$. In the no-molecule reference, all spin-current components remain strongly suppressed across the full $\alpha$ range, consistent with the absence of spin-active scattering. Once the molecular field is present, finite spin currents develop and grow monotonically with the effective SOC strength $\alpha_{\mathrm{SO}}$. In the superconducting regime these currents acquire a phase-coherent Josephson component originating from the spin structure of the Andreev states, and the two enantiomers separate clearly in each spin channel [Figs.~\ref{fig:alpha}(b)--\ref{fig:alpha}(d)]. 

The separation between enantiomers in each spin channel originates from the molecular spin--orbit coupling, and the sign structure of the three components together with the difference in their magnitudes follows from the molecular mirror symmetry and the anisotropic form of the potential. We analyze this in Sec.~\ref{subsec:theta-scan}.
Overall, Fig.~\ref{fig:alpha} identifies $\alpha$ as a practical control knob for amplifying spin-polarized Josephson signatures of chirality. This motivates using spin-resolved observables (and, where appropriate, additional symmetry-breaking fields) as the primary experimental pathway to robust handedness detection in superconducting weak links.

\subsection{Angular dependence and symmetry constraints}\label{subsec:theta-scan}

\begin{figure}
    \centering
    \includegraphics[width=\columnwidth]{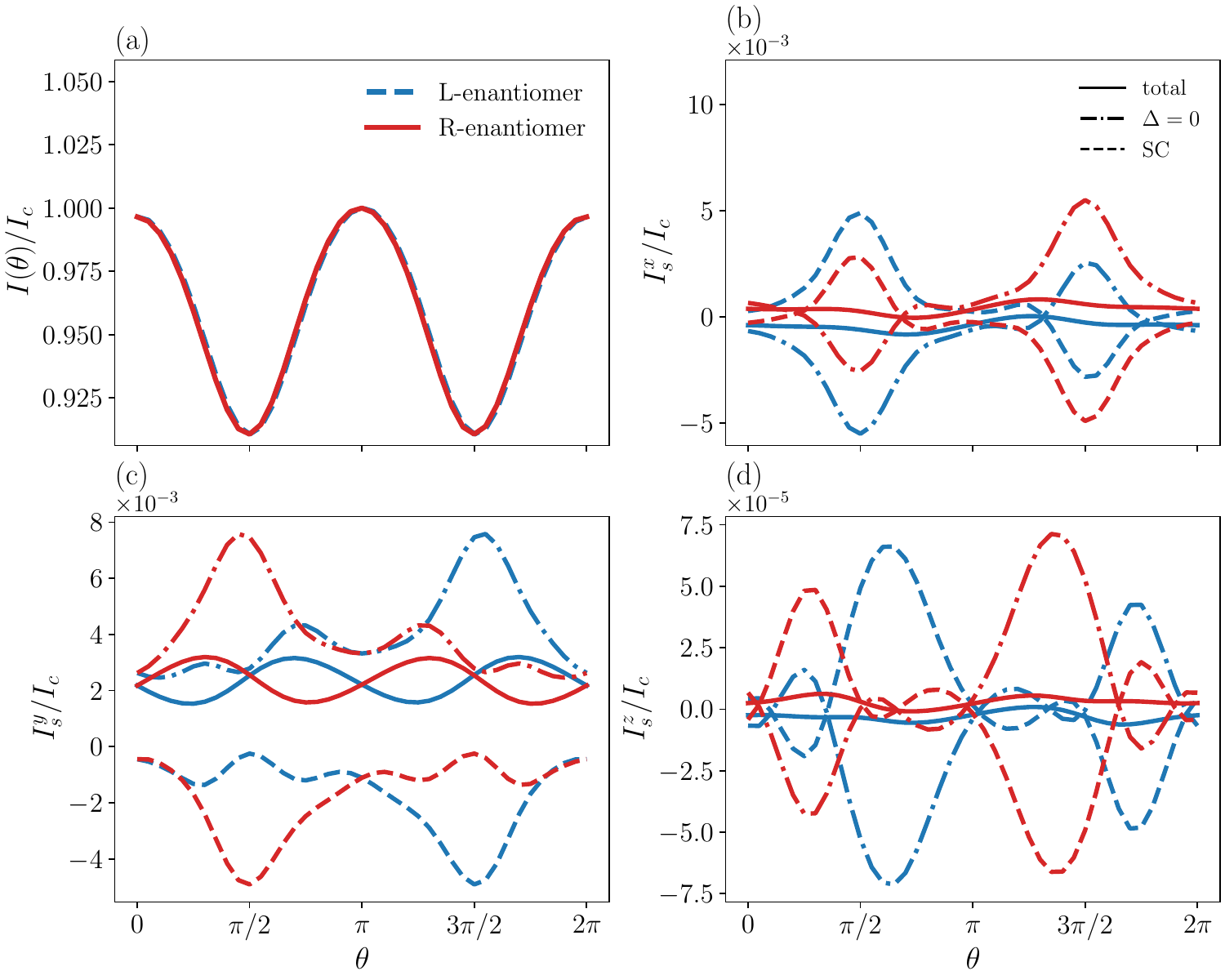}
    \caption{
Angular dependence of the charge and spin currents on the molecular tilt angle $\theta$. 
Rotating the molecular electrostatic texture modifies the effective spin--orbit field experienced 
by quasiparticles in the junction. 
While the charge current remains only weakly sensitive to chirality, the spin-current components
show a pronounced orientation dependence and distinct chirality-dependent responses for the two enantiomers.
As in Fig.~\ref{fig:alpha}, the total current (solid) is separated into the normal-state part at $\Delta=0$ (dash-dotted) and the superconductivity-induced part (dashed), for the left-handed (blue) and right-handed (red) enantiomers, all normalized to $I_c$. The angle ticks are in radians.
Parameters: $W=16$, $L=46$, $\mu/t=0.5$, $\Delta_0/t=0.2$, $\alpha_{\mathrm{SO}}/t=2$, $h_z=0$, $\phi=\pi/2$, and $B=0$.
}
    \label{fig:theta}
\end{figure}

In fig.~\ref{fig:theta}, we show how the Josephson response varies with the molecular tilt angle $\theta$, which rotates the molecular electrostatic texture relative to the junction plane and thereby changes the structure of the effective spin--orbit field seen by traversing quasiparticles~\cite{Moharana2025}. The curves are obtained by repeating the equilibrium calculations described above while varying $\theta$ at fixed superconducting phase bias and fixed $\alpha$. Formally, $\theta$ modifies the spatial gradients of the scalar potential $V(\mathbf{r})$ and hence the effective spin--orbit term of the generic form $(\nabla V\times \mathbf{p})\cdot \boldsymbol{\sigma}$, which in turn alters the spin structure of the Andreev eigenstates entering Eqs.~\eqref{eq:Josephson_from_spectrum_final} and \eqref{eq:spin-current}.

The effect of chirality is more accentuated in fig.~\ref{fig:theta} where chirality detection can be optimized by orientational control. The structure of these curves follows from the mirror $M_{xz}$ that relates the two enantiomers, the reflection $y\to-y$. Since $\theta$ rotates the molecule about $\hat{x}$, this mirror reverses it as well, acting as $(\chi,\theta)\to(-\chi,-\theta)$. The charge current along $\hat{x}$ lies in the mirror plane and is even, while spin is a pseudovector whose in-plane components flip, so $\sigma_x$ and $\sigma_z$ are odd and $\sigma_y$ is even. Hence $I_s^x$ and $I_s^z$ are odd under this full mirror and $I_s^y$ is even, where even and odd refer to $(\chi,\theta)\to(-\chi,-\theta)$ and not to reversing $\chi$ at fixed $\theta$. This is transparent in Eq.~\eqref{eq:mol_soc_compact}, where on a bond along $\hat{x}$ the spin--orbit coupling reduces to $\mathcal{E}_z\,\sigma_y-\mathcal{E}_y\,\sigma_z$ and the chirality-odd dipole $\mu_y$ enters through $\mathcal{E}_y$. The component $I_s^y$ instead couples to $\mathcal{E}_z$, which carries no handedness at $\theta=0$ and becomes chirality dependent only once a finite tilt mixes $\mu_y$ into it. The figures compare $\chi$ and $-\chi$ at fixed $\theta$, the experimentally relevant orientation, so the enantiomer curves are not pointwise mirror images and their magnitudes and apparent periodicities are set by the anisotropic potential. The comparison between $\chi=\pm1$ shows that the chirality-odd response persists across a broad angular range, while specific angles can maximize the contrast between enantiomers. This angular tunability is particularly relevant experimentally because it suggests that geometric alignment (or controlled adsorption orientation) can be used to amplify chirality-sensitive observables without changing material parameters.

\subsection{Temperature dependence and experimental feasibility}\label{subsec:T-scan}

\begin{figure}
    \centering
    \includegraphics[width=\columnwidth]{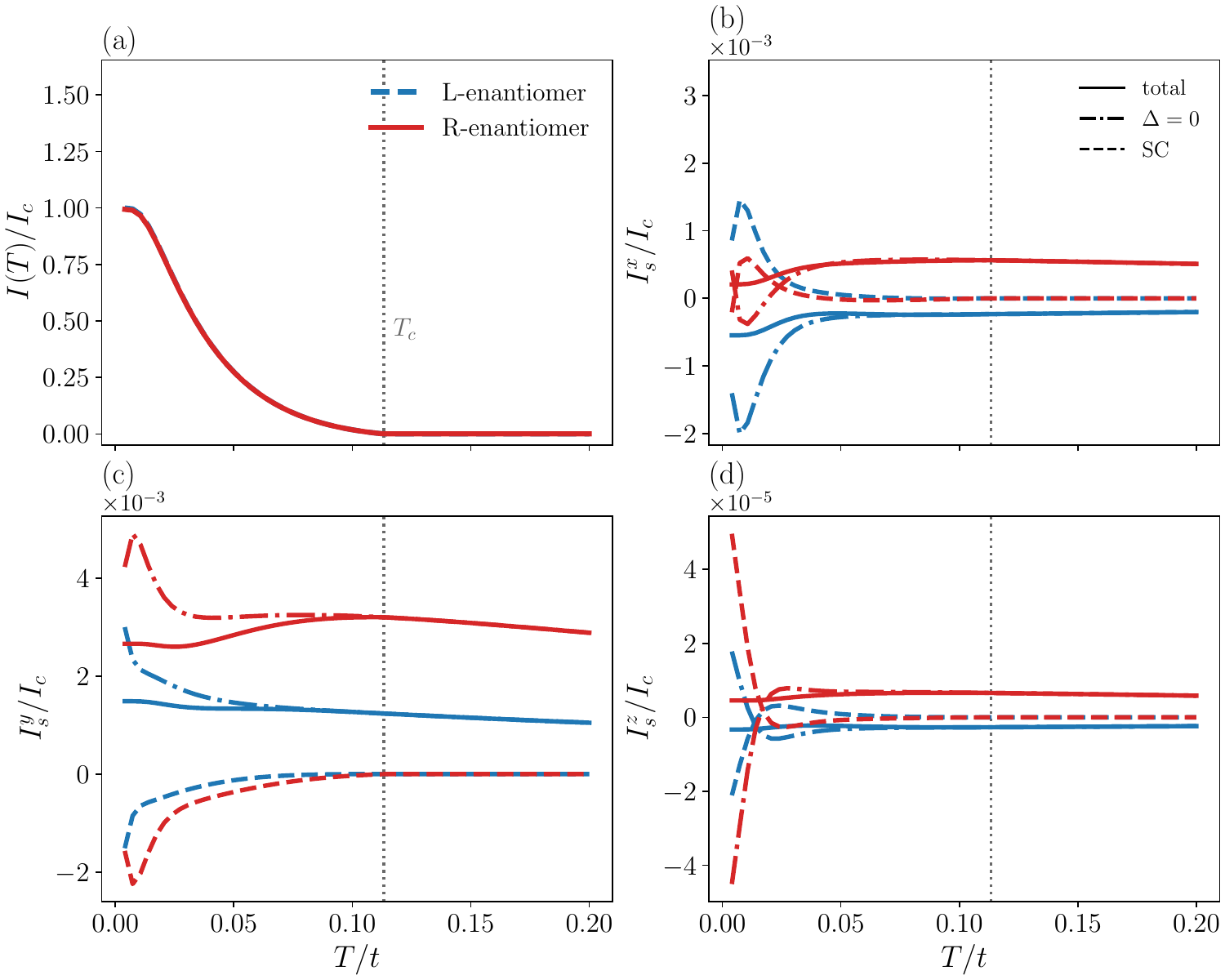}  
\caption{
Temperature dependence of charge and spin currents in the chiral-molecule Josephson junction. 
(a) The normalized charge current decreases with increasing temperature and vanishes as the 
superconducting gap closes near $T_c$. 
(b)--(d) The spin-current components are progressively modified with temperature but retain a residual offset above $T_c$.
This offset reflects an equilibrium spin-current background generated by the spin--orbit-coupled
molecular potential rather than a phase-coherent Josephson contribution.
The decomposition makes this explicit. The total current (solid) is the sum of the normal-state part at $\Delta=0$ (dash-dotted), which persists above $T_c$, and the superconductivity-induced part (dashed), which vanishes at $T_c$, shown for the left-handed (blue) and right-handed (red) enantiomers and normalized to $I_c$. The vertical line marks $T_c$.
Parameters: $W=16$, $L=46$, $\mu/t=0.5$, $\Delta_0/t=0.2$, $\alpha_{\mathrm{SO}}/t=2$, $\theta=\pi/4$,
$\phi=3\pi/4$, and $B=0$.
}
    \label{fig:T}
\end{figure}
In Fig.~\ref{fig:T} we test the robustness of the chiral signatures against thermal effects. Temperature enters the equilibrium currents in two complementary ways. First, it controls quasiparticle occupations through the factors $\tanh(E_n/2T)$ in Eqs.~\eqref{eq:Josephson_from_spectrum_final} and \eqref{eq:spin-current}, progressively reducing the contribution of higher-energy states and smoothing the phase sensitivity of the spectrum. Second, it suppresses the superconducting order parameter in the leads, which directly reduces the Andreev level dispersion and hence the spectral slopes that control both charge and spin supercurrents. We incorporate this second effect by taking the lead gap $\Delta(T)$ from the Bardeen-Cooper-Schrieffer (BCS) temperature dependence.

The resulting trends in Fig.~\ref{fig:T} are consistent with this physical picture. The charge Josephson response decreases monotonically with $T$ and is strongly suppressed as the gap $\Delta(T)$ closes and the presence of the chiral molecule decreases the current.
In contrast, the spin-current components exhibit two distinct contributions. The superconductivity-induced component decreases with increasing temperature in parallel with the suppression of the superconducting proximity effect. However, a residual signal remains even when $T\gtrsim T_c$. This offset originates from the spin–orbit-coupled molecular texture and reflects an equilibrium spin-current background of the normal-state Hamiltonian rather than a Josephson contribution. The decomposition into normal-state and superconducting contributions defined in Eq.~\eqref{eq:sc-spin-current} is resolved for each spin component in Fig.~\ref{fig:T}. Such equilibrium spin currents are known to arise in systems with broken inversion symmetry and spin–orbit coupling when the conventional spin-current operator is used~\cite{yi2006a,sun2007}. Therefore, the superconductivity-enabled contribution can be identified operationally as the component that scales with the superconducting gap $\Delta(T)$, whereas the total chirality-dependent signal contains both a normal-state and a superconducting contribution. Both contributions are chirality odd, but they have different microscopic origins and different temperature dependences. The normal-state part is the equilibrium spin current of the broken-inversion spin-orbit-coupled normal state and persists above $T_c$, while the superconducting part originates from the chirality-dependent modification of the Andreev spectrum and vanishes together with $\Delta(T)$. The genuinely Josephson-specific enantiomer contrast is therefore the gap-scaling component $I^a_{s,\mathrm{SC}}$.

\section{Conclusion}
We have shown that chiral molecular electrostatic textures, when integrated into an SNS Josephson junction, act as a tunable source of spin-dependent scattering that restructures the Andreev bound-state spectrum and thereby modifies equilibrium supercurrents. Across the explored parameter space, the most robust chirality-sensitive signatures arise in spin-polarized equilibrium transport: opposite enantiomers generate systematically different spin-polarized currents, with a pronounced anisotropy among spin components that reflects the nontrivial spatial structure of the molecular potential and its associated SOC field. By contrast, the equilibrium charge current remains nearly identical in symmetric, zero-field configurations, consistent with chirality entering primarily through the spin texture of quasiparticle states while the total charge response is constrained by equilibrium symmetries. Importantly, the spin response is not only chirality-dependent but also tunable, where varying the effective SOC strength and molecular orientation provides practical control knobs that amplify the handedness contrast without requiring changes to the junction geometry. Our results complement recent efforts to engineer quantum transport
phenomena through symmetry control in condensed matter systems,
including recent demonstrations of nonreciprocal superconducting
transport in structurally chiral Josephson junctions~\cite{orban2026}.
Adding a small time-reversal-breaking field to the chiral weak link studied here is a natural route to a chirality-controlled Josephson diode, because the molecular spin-orbit texture already supplies the inversion breaking and the spin-active scattering that the large chirality-odd spin response reflects~\cite{Debnath2024,Costa2025}. We stress, however, that broken inversion and broken time-reversal symmetry are necessary but not sufficient for a diode response, so the magnitude of such an effect is set by the resulting asymmetry of the Andreev spectrum rather than by the equilibrium spin-current magnitude alone, and we leave its quantitative study to future work.

From an application standpoint, these results motivate a phase-sensitive strategy for probing chirality-dependent spin structure using superconducting circuits. 
Chirality can in general be detected through spin-selective transport processes. However, the Josephson platform probes the phase accumulated by quasiparticles traversing the chiral molecular potential. 
The resulting interferometric response therefore provides direct access to the coherence of the underlying spin-dependent scattering processes. 
The present results therefore identify which spin-polarized observables are most sensitive to molecular handedness and clarify why purely charge-based Fraunhofer signals remain weakly enantiomer-dependent in the symmetric zero-field geometry considered here.

\begin{acknowledgments}
O.~K. is funded by the European Union, in the framework of the Marie Skłodowska-Curie Actions- CISSE Doctoral Network, under Grant Agreement No. 101071886. Views and opinions expressed are, however, those of the author(s) only and do not necessarily reflect those of the European Union or the European Research Executive Agency (REA). Neither the European Union nor the granting authority can be held responsible for them.
R.A. received funding from the Austrian Academy of Sciences (ÖAW), Grant No. PR1029OEAW03.
\end{acknowledgments}
\appendix
\setcounter{figure}{0}
\renewcommand{\thefigure}{A\arabic{figure}}
\setcounter{table}{0}
\renewcommand{\thetable}{A\arabic{table}}
\section{Molecular potential and simulation parameters}
\label{appA:mol_potential}
In the numerical simulations presented in this work the chiral molecule is modeled through an electrostatic potential acting on the lattice sites of the normal region. The functional form of this potential follows the parametrization introduced in Refs.~\cite{Ghazaryan2020a,alhyder2025b} and reproduced here for completeness.

The molecular electrostatic potential is written as

\begin{equation}
V_{\mathrm{mol}}(\mathbf r)
=
\mathbf \mathcal{E} \cdot \mathbf r'
\exp\!\left[
-\frac12
\left(
\frac{x'^2}{\ell_x^2}
+
\frac{y'^2}{\ell_y^2}
+
\frac{z'^2}{\ell_z^2}
\right)
\right],
\end{equation}

where $\mathbf r'=(x',y',z')$ denotes coordinates in the molecular frame. These coordinates are obtained from the laboratory frame position $\mathbf r=(x,y,z)$ through a rotation around the $x$ axis by an angle $\theta$,
$
\mathbf r'
=
R_x(\theta)\,
(\mathbf r-\mathbf r_m),
$
with $\mathbf r_m=(x_m,y_m,0)$ the molecular center and $R_x(\theta)$ the standard rotation matrix.

The parameters $\ell_x$, $\ell_y$, and $\ell_z$ control the spatial extent of the potential along the three molecular axes. The linear prefactor is determined by an effective internal electric field $\mathbf{\mathcal{E}}=(\mathcal{E}_x,\mathcal{E}_y,\mathcal{E}_z)$ related to a molecular dipole moment $\boldsymbol{\mu}$ through
$
\mathcal{E}_\nu = \frac{8 e\, \mu_\nu}{\ell_\nu^3},
$ with $
\nu\in\{x,y,z\}.
$
The Nambu basis ensures electrons and holes experience opposite scattering potentials.

Opposite molecular enantiomers are implemented by reversing the $y$ component of the molecular dipole moment,
$
\mu_y \rightarrow \chi\, \mu_y ,
$ with $
\chi = \pm 1 ,$

while all other parameters are kept fixed. This transformation corresponds to a mirror operation with respect to the $xz$ plane in the molecular frame and reverses the handedness of the electrostatic texture.

Unless stated otherwise, all numerical results in this work use the same molecular parameters:
$
\ell_x = 2.0, 
\ell_y = 4.0, 
\ell_z = 10.0,
$
and dipole moments
$
\mu_x = 2.4,
\mu_y = 5.0 \chi,
\mu_z = 1.8.
$
The molecular orientation angle is fixed to $
\theta = \pi/4,
$
unless explicitly varied in the angular dependence calculations. 
We also use an overall dimensionless scale factor in the molecular electrostatic potential so that the molecular perturbation remains smaller than the kinetic bandwidth. Changing this scale modifies the absolute magnitude of the spin-polarized response, but the symmetry relations and the qualitative chirality contrast are unchanged as long as the molecular potential remains perturbative. In physical units, taking the hopping $t\approx 1.14$ eV set by the effective mass and lattice spacing, the peak molecular potential $\max|V_{\mathrm{mol}}|$ is about $0.11$ eV at the reference scale and $0.55$ eV at the strongest scale used in the figures, and the induced spin--orbit hopping $\alpha_{\mathrm{SO}}\,|\nabla V_{\mathrm{mol}}|$ is of order $50$ to $270$ meV, consistent with the effective spin--orbit scales inferred for chirality-induced spin selectivity~\cite{Ghazaryan2020a}. The superconducting gap is fixed at $\Delta=0.2\,t$, with the Fermi energy $E_F=0.5\,t$ measured from the band bottom.

The molecule is positioned at the center of the junction,
and the strength of the induced spin–orbit coupling is controlled by the dimensionless parameter $\alpha_{\mathrm{SO}}$ which can be varied. The molecular electrostatic profile is not confined to the weak link but extends across the junction, consistent with an adsorbed sheet of chiral molecules covering the device. We treat $\alpha_{\mathrm{SO}}$ as an effective phenomenological coupling that accounts for microscopic renormalization of the Pauli spin--orbit term by the molecular environment, interface fields, and substrate screening. It should therefore be read as a control parameter for the induced spin-active scattering rather than as a bare atomic spin--orbit constant. The parameter is dimensionless and multiplies the field gradient, so the associated spin--orbit energy is $\alpha_{\mathrm{SO}}$ times that gradient and stays in the tens of meV range across the values used here, rather than of order $t$.

The equilibrium currents are obtained by full diagonalization of the finite Bogoliubov--de Gennes Hamiltonian, so the spectral sum is complete and no level truncation or branch tracking is involved. The molecular spin--orbit term is discretized symmetrically on each bond, evaluated at the bond midpoint, which makes the mirror operation $M_{xz}$ an exact lattice symmetry and renders the charge currents of mirror-related configurations $(\chi,\theta)$ and $(-\chi,-\theta)$ identical to machine precision. The spin currents reported here are the sum of bond currents across a fixed transverse cut at the center of the normal region. Because spin is not conserved in the presence of spin--orbit coupling, this conventional spin current is not a conserved transport current, and its flux through a transverse cut depends on the position of the cut. We therefore interpret it as a local equilibrium response evaluated at a stated cut rather than as a cut-independent transport observable. The chirality-dependent response is controlled by symmetry and by the junction geometry rather than by numerical truncation. The mirror parities of the charge and spin currents hold to machine precision at every lattice size we diagonalize, the response vanishes to machine precision when the molecular potential is removed and grows monotonically with $\alpha_{\mathrm{SO}}$, and the sign and structure of the enantiomer contrast are stable across sizes. The currents converge with the lead length $L$, saturating through damped finite-size oscillations once the leads exceed the coherence length, whereas the width $W$ sets the number of transverse channels and thus the device size. We therefore fix a lead length in the converged regime and report the currents as representative of a junction of the chosen width.

\section{Current operators and numerical evaluation}
\label{appB:current_operators}
The form of Eq.~\ref{eq:spin-current} follows directly from the thermal
expectation value of a one-body operator in the Bogoliubov--de
Gennes basis. Let $\hat{\mathcal J}^{a}$ denote the many-body
operator associated with the spin-current matrix $\hat J^{a}$, with $a\in\{x,y,z\}$ the spin polarization direction. In a lattice formulation the current is
naturally defined on bonds between neighboring sites.
Starting from the physical BdG spin-density matrix
$\mathcal S_a=\mathrm{diag}(\sigma_a,-\sigma_a^\ast)$,
the conventional spin-polarized bond-current operator can be written
for a bond $(i,j)$ as
\begin{equation}
\hat J^a_{ij}
=
\frac{i}{2}
\left(
\Psi_i^\dagger
\mathcal S_a H_{ij}
\Psi_j
-
\Psi_j^\dagger
H_{ji}
\mathcal S_a
\Psi_i
\right),
\end{equation}
where $H_{ij}$ denotes the Bogoliubov--de Gennes
Hamiltonian matrix element connecting sites $i$ and $j$
(including both kinetic and spin--orbit hopping terms).
The total spin current across a transverse cut of the
junction is obtained by summing $\hat J^a_{ij}$ over all
bonds crossing that cut. Because the molecular spin--orbit interaction breaks spin conservation, this bond current is not a conserved current in the same sense as charge current. The nonconservation appears as local spin torque terms, and the quantity computed here should therefore be interpreted as the conventional equilibrium spin-current response.
In practice, after diagonalization of $\hat H_{\mathrm{BdG}}$, the quasiparticle
spectrum appears in particle--hole-symmetric pairs
$\pm E_n$, with $E_n>0$, and the equilibrium occupation of a
quasiparticle mode is given by the Fermi function
$
f(E_n)=1/(e^{E_n/T}+1),
$
where we use $k_B\equiv 1$ throughout. In the BdG formalism,
the expectation value of any bilinear operator can be written as
a sum over the contributions of the $\pm E_n$ partners. For the
spin-current operator this gives
\begin{equation}
I_s^{a}
=
\frac{1}{2}\sum_{n>0}
\left[
f(E_n)\,\langle n|\hat J^{a}|n\rangle
+
\bigl(1-f(E_n)\bigr)\,
\langle \bar n|\hat J^{a}|\bar n\rangle
\right],
\end{equation}
where $|\bar n\rangle$ is the particle--hole partner of $|n\rangle$
with energy $-E_n$. The prefactor $1/2$ removes the double
counting inherent to the Nambu representation.

Using particle--hole symmetry, the matrix elements of the
current operator for the two partner states are related by
$
\langle \bar n|\hat J^{a}|\bar n\rangle
=
-\langle n|\hat J^{a}|n\rangle,
$
so that the two terms combine into
$$
I_s^{a}
=
\frac{1}{2}\sum_{n>0}
\left[
1-2f(E_n)
\right]
\langle n|\hat J^{a}|n\rangle.
$$
Using the identity
$
1-2f(E)=\tanh\!\left(\frac{E}{2T}\right),
$
one obtains Eq.~\ref{eq:spin-current},
\begin{equation}
I_s^{a}
=
\frac{1}{2}\sum_{n>0}
\tanh\!\left(\frac{E_n}{2T}\right)
\langle n|\hat J^{a}|n\rangle.
\end{equation}
\begin{figure}[t]
\centering
\includegraphics[width=0.9\linewidth]{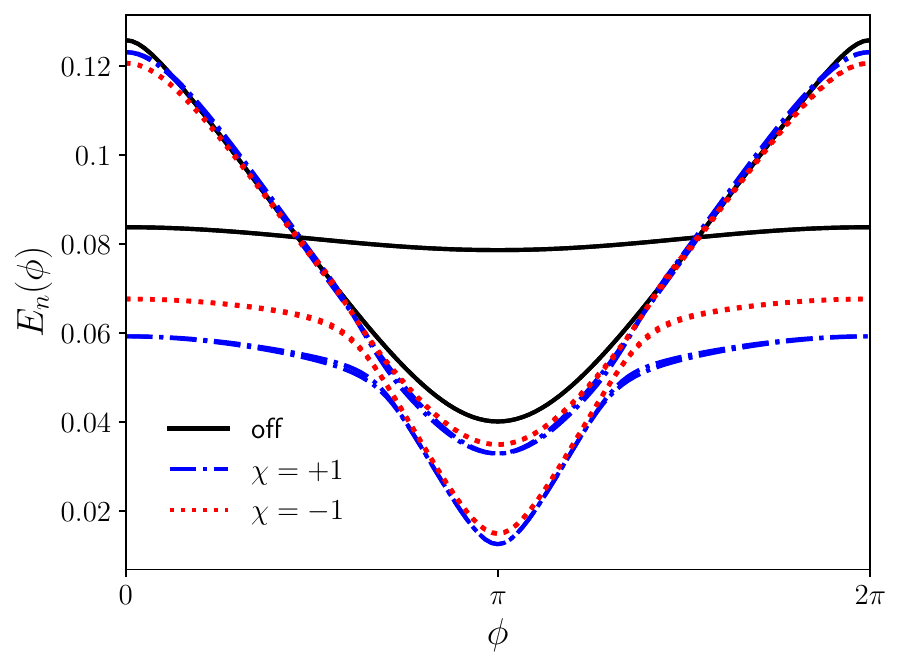}
\caption{Low-energy Andreev bound-state branches $E_n(\phi)$ obtained from diagonalization of the Bogoliubov–de Gennes Hamiltonian. Solid lines correspond to the junction without molecules, while dashed and dotted lines show the spectra for molecular potentials with opposite chiralities $\chi=\pm1$.
}
\label{fig:abs_appendix}
\end{figure}
This representation is convenient because the sum can be
restricted to positive BdG energies only, while the thermal
weight $\tanh(E_n/2T)$ compactly accounts for the occupations
of both members of each particle--hole pair. Equivalently, one
may write the thermal factor as $1-2f(E_n)$, but the
$\tanh(E_n/2T)$ form is more compact and is standard in BdG
expressions for equilibrium observables.
\section{Andreev bound-state spectrum and phase dispersion}
\label{appD:abs_spectrum}
To illustrate the spectral origin of the equilibrium Josephson response, we compute the low-energy positive Bogoliubov--de Gennes eigenvalues as a function of the superconducting phase difference $\phi$. For each value of $\phi$, the finite-system BdG Hamiltonian is diagonalized numerically and the eigenvalues closest to zero energy are extracted. The corresponding branches are then tracked between neighboring phase points by maximizing the overlap between eigenvectors, which yields a continuous representation of the low-energy phase-dependent spectrum.

Figure~\ref{fig:abs_appendix} shows the resulting low-energy branches for the junction without molecules and for junctions with the molecular potential present. These branches form the subgap sector responsible for the phase-sensitive Josephson response. In particular, the charge current is governed by the phase derivatives of the BdG energies (See Eq.~\eqref{eq:Josephson_from_spectrum_final}),
so that changes in the spectral dispersion directly translate into changes in the current--phase relation. At low temperature, these low-energy Andreev branches dominate the Josephson current.

The comparison shows that the molecular potential modifies the low-energy phase dispersion relative to the reference junction without molecules, consistent with the molecule acting as a spin-dependent scatterer in the normal region. At the same time, the spectra for opposite enantiomers remain rather similar in this symmetric zero-field configuration, in agreement with the weak chirality dependence of the equilibrium charge current discussed in the main text.
\bibliography{references}

\end{document}